\documentclass[aps,twocolumn,showpacs,superscriptaddress,preprintnumbers,amsmath,amssymb, pra]{revtex4}
\usepackage[dvips]{graphicx}
\usepackage{dcolumn}
\usepackage{color}
\usepackage{bm}

\newcommand{\TKKcomp}{Department of Applied Physics/COMP, Aalto University, P.O.~Box 15100, FI-00076 AALTO, Finland}
\newcommand{\TKKltt}{Low Temperature Laboratory, Aalto University, 
  P.O.~Box 13500, FI-00076 AALTO, Finland.}
\newcommand{\CQCT}{Australian Research Council Centre of Excellence for Quantum Computer
    Technology, School of Electrical Engineering \& Telecommunications,
    University of New South Wales, Sydney, New South Wales 2052, Australia.}

\begin{document}
\title{Ground-state geometric quantum computing in superconducting systems}

\author{P. Solinas}
\affiliation{\TKKcomp}

\author{ J.-M. Pirkkalainen}
\affiliation{\TKKcomp}
\affiliation{\TKKltt}

\author{M. M\"ott\"onen}
\affiliation{\TKKcomp}
\affiliation{\TKKltt}
\affiliation{\CQCT}

\begin{abstract}
We present a theoretical proposal for the implementation of geometric quantum computing based on a Hamiltonian which has a doubly degenerate ground state. 
Thus the system which is steered adiabatically, remains in the ground-state.
The proposed physical implementation relies on a superconducting circuit composed of three SQUIDs and two superconducting islands with the charge states encoding the logical states.
We obtain a universal set of single-qubit gates and implement a non-trivial two-qubit gate exploiting the mutual inductance between two neighboring circuits,  allowing us to realize a fully geometric ground-state quantum computing.
The introduced paradigm for the implementation of geometric quantum computing is expected to be robust against environmental effects.
\end{abstract}

\pacs{03.67.Lx, 03.65.Vf, 85.25.Dq}

\maketitle 

\section{ Introduction}
One of the major issues in the present quantum information technology is the protection of quantum systems from external noise.
The noise can be due to imprecise control or undesired interaction with the environment.
To eliminate some errors of the first kind, geometric quantum computing (GQC) has been proposed \cite{zanardi99, jones00,falci00,ekert00}.
The idea behind it is to use the so-called geometric phases \cite{berry84} to manipulate the quantum state.
Usually the appearance of a geometric phase is associated with a cyclic and adiabatic manipulation of the system.
The advantage is twofold: the adiabatic evolution eliminates the problem of fine-tuning of the evolution time and the geometric operators can be intrinsically robust against fluctuations of the imprecise control field \cite{dechiara03}.

It is not possible to realize universal quantum computing only by means of the Abelian Berry phase \cite{berry84}. Therefore, utilizing the non-Abelian geometric phases \cite{wilczek84}, the geometric quantum computing was proposed in 1999 \cite{zanardi99}.
In this case, the logical states are encoded in a degenerate subspace of the full Hilbert space and they can be manipulated by means of purely geometric non-Abelian operators.
The first proposals of the physical implementations were based on trapped ions and exploited the so-called tripod Hamiltonian to build a doubly degenerate logical subspace \cite{unanyan99, duan01}.
Subsequently, the same Hamiltonian has been used in many quantum systems providing means for possible implementation for neutral atoms, superconducting circuits, quantum dots, and Bose-Einstein condensates \cite{fuentes02,recati02,faoro03,solinas03,zangh05}.
In conjunction with these proposals, various studies have been carried out to test the stability of geometric gates against different types of noise. Here, GQC has proven robust against fast adiabatic fluctuations of control fields \cite{solinas04} and certain type of environmental noise \cite{parodi07, florio06}.

Despite the theoretical efforts, we are still far from a full GQC, the experimental observations of non-Abelian phases remain very limited and still under discussion \cite{morton2005}.
One possible reason behind the absence of experimental verification is the decoherence due to the interaction of the quantum system with the environment.
In fact, because of the condition of adiabatic evolution, it is challenging to protect the system for long enough time to implement geometric operations.
This is the case for all of the above proposals based on tripod-like Hamiltonians in which the evolution does not occur in the lowest-energy eigenspace.
In general, one expects that decoherence such as relaxation can be avoided if the evolution occurs in the ground-state manifold. 
The possibility to exploit this robustness in large systems to perform adiabatic quantum computation is still under discussion \cite{ashhab06}.
However, the robustness of ground state evolution has been recently confirmed for a driven two-level system in the adiabatic limit~\cite{pekola09,solinas10,salmilehto10}.
In the same studies, it was shown that relaxation induced by the environment can help to keep the system in the ground state potentially leading to an improved robustness of the ground-state evolution.

These results suggest that GQC which exploits a degenerate ground-state eigenspace can be experimentally feasible.
The first step in this direction has been taken in Ref.~\cite{pirkkalainen10} in which a way to detect non-Abelian phases in superconducting Josephson devices was proposed.
However, the work mainly focused on observing the geometric non-Abelian effects and only a particular unitary transformation was discussed. 

The extension to a universal set of transformations is non-trivial and is the main result of this paper.
Starting from a general Hamiltonian, we show how to obtain the basic logical gates in an abstract context.
Then, using the same system as in Ref.~\cite{pirkkalainen10}, we show how these gates can be implemented in a physical set-up.
We build all the logical gates needed for quantum computing: a complete set of single-qubit gates and a two-qubit gate exploiting the mutual inductance between two circuits next to each other.
Together these gates form an universal set allowing full geometric quantum computing.

The paper is organized as follows. In Sec.~\ref{sec:model}, we introduce a model Hamiltonian with a doubly degenerate ground state which can produce universal single-qubit transformations.
In Sec.~\ref{sec:Hphys}, we present the physical system and show how to map the abstract model into the physical Hamiltonian.
In Sec.~\ref{sec:single_qubit}, we show how to manipulate the quantum system in order to obtain two fundamental single-qubit gates.
Section~\ref{sec:two_qubit} is devoted to the implementation of a geometric two-qubit gate.
In Sec.~\ref{sec:measurement}, the readout measure is discussed and Sec.~\ref{sec:conclusions}  concludes the paper.

\section{Model}
\label{sec:model}

We consider a quantum system with a two-fold degenerate ground-state eigenspace and an excited state.
The system is steered adiabatically along a loop by external control fields.
If the evolution is adiabatic and occurring in time $T_{ad}$, the dynamics can be accurately determined by instantaneous diagonalization of the system Hamiltonian since the system follows approximately the evolution of the eigenstates.
It is convenient to fix the logical basis in order to have the initial Hamiltonian in a diagonal form: in particular, the two degenerate ground-states used as logical states are denoted as $|0\rangle$ and $|1\rangle$, and the excited one representing an auxiliary state is denoted as $|a\rangle$.
Taking the ground and excited state energies equal to $ \eta$ and $\eta+\alpha$, respectively, we have
$\hat{H}(t=0)=  \eta \openone + \alpha |a\rangle \langle a|$, where $\openone= |0\rangle \langle 0|+|1\rangle \langle 1|+|a\rangle \langle a|$.
The basis  $\{ |a\rangle , |0\rangle, |1\rangle \}$ is time-independent and corresponds to the eigenbasis only if the Hamiltonian is in the above form.

We require that the Hamiltonian has always a doubly degenerate ground state during the control cycle.
Thus we can formally write it as $\hat{H}(t)=  \eta \openone +\alpha |v_1\rangle \langle v_1|$ where $|v_1\rangle$ is a linear combination of the time-independent $\{ |a\rangle , |0\rangle, |1\rangle \}$ states.
Throughout the paper we adopt the convention to denote with the operators with hat symbol (e.g., $\hat{H}$) and their representation in the basis $\{ |a\rangle , |0\rangle, |1\rangle \}$ with the tilde (e.g., $\tilde{H}$).
Following Ref.~\cite{niskanen02}, the most general representation with the spectrum of $\hat{H}$ in the basis  $\{ |a\rangle , |0\rangle, |1\rangle \}$ reads 
\begin{widetext}
\begin{equation}
\tilde{H}=  \left(\begin{array}{lll}
  \eta+\alpha \cos ^2\theta _1\ \cos ^2\theta_2 & - \alpha e^{-i \phi _1} \cos
   \theta_1 \cos ^2\theta_2 \sin \theta_1 &
   -\alpha e^{-i \phi _2} \cos \theta_1 \cos \theta_2 \sin \theta_2 \\
 -\alpha e^{i \phi _1} \cos \theta_1 \cos ^2 \theta_2 \sin \theta_1 &  \eta + \alpha \cos ^2\theta_2 \sin ^2\theta _1
   & \alpha e^{i (\phi _1-\phi_2)} \cos \theta_2 \sin \theta_1 \sin \theta_2 \\
 -\alpha e^{i \phi _2} \cos \theta_1 \cos \theta_2 \sin
   \theta_2 & \alpha e^{-i (\phi _1-\phi_2)} \cos \theta
  _2 \sin \theta_1 \sin \theta_2 &  \eta+ \alpha \sin^2\theta_2
\end{array}
\right),
\label{eq:H_general}
\end{equation}
\end{widetext}
where $\phi_1$, $\phi _2$, $\theta_1$, and $ \theta_2$ are experimentally modulated parameters.
In the following we will assume that $\phi _1=0$.
The parameters  $ \theta_1$, $ \theta_2$, and $ \phi_2$ describe effectively the time evolution of the Hamiltonian.
The instantaneous eigenvectors of $\hat{H}$ can be written as  
\begin{eqnarray}
  |v_1 \rangle &=& -e^{-i \phi_2} \cos \theta_1 \cos \theta_2 |a\rangle +e^{-i \phi_2}\cos \theta_2 \sin \theta_1 |0\rangle \nonumber \\  
  &&+  \sin \theta_2 |1\rangle ,  \nonumber \\  
  | v_2 \rangle &=& \sin \theta_1 |a \rangle +\cos \theta_1 |0 \rangle \nonumber,  \\  
   | v_3 \rangle&=& e^{-i \phi_2}\cos \theta_1 \sin \theta_2 |a \rangle  - e^{-i \phi_2}\sin \theta_1 \sin \theta_2 |0 \rangle\nonumber \\
   &&+ \cos \theta_2 |1 \rangle,
     \label{eq:eigenvectors}
\end{eqnarray}
where $| v_2 \rangle$ and $| v_3 \rangle$ are the degenerate time-dependent ground states.

The restriction to adiabatic evolution, i.e., $\alpha T_{ad}/ \hbar \gg 1$, assures that no transitions to the excited state occurs; the evolution is restricted to the ground-state eigenspace. 
The final unitary transformation depends only on the geometric features, i.e.,  on the loop $\gamma$ covered in the parameter space $\{ \theta_1, \theta_2 , \phi_2\}$, and it reads  \cite{wilczek84,zanardi99,niskanen02}
\begin{equation}
  \hat{U}_{\gamma} =\mathcal{P} \mathrm{exp} \Big [- \oint_\gamma (\hat{A}_{\theta_1} d \theta_1+\hat{A}_{\theta_2} d \theta_2+\hat{A}_{ \phi_2} d  \phi_2) \Big],
\end{equation}
where $\mathcal{P}$ denotes the path ordering operator.
The $\hat{A}_{x} $ are the so-called connections and, restricted to the degenerate eigenspace, they are defined by $\tilde{A}_{x,j-1,i-1}= \langle v_j| \partial_x |v_i\rangle$ ($i,j=2,3$).
In our case, restricted to the degenerate eigenspace, they assume a simple form
\begin{eqnarray}
\tilde{A}_{\theta_1}&=&\left(
\begin{array}{ll}
 0 & e^{i \phi_2} \sin \theta_2 \\
 -e^{-i \phi_2} \sin \theta_2 & 0
\end{array}
\right),
 \nonumber \\
\tilde{A}_{\theta_2}&=&
\left(
\begin{array}{ll}
 0 & 0 \\
 0 & 0
\end{array}
\right),
 \nonumber \\
\tilde{A}_{\phi_2}&=&
\left(
\begin{array}{ll}
 0 & 0 \\
 0 & - i \sin^2 \theta_2
\end{array}
\right).
\label{eq:connections}
\end{eqnarray}
Since $\hat{A}_{\theta_2}$ gives no contribution,
the unitary transformation associated with a loop $\gamma$ in the parameter space $\{ \theta_1, \theta_2 , \phi_2\}$ can be written as 
\begin{equation}
  \hat{U}_{\gamma} =\mathcal{P} \mathrm{exp} \Big [- \oint_\gamma (\hat{A}_{\theta_1} d \theta_1 +\hat{A}_{ \phi_2} d  \phi_2) \Big].
  \label{eq:U_gamma}
\end{equation}

\begin{figure*}
    \begin{center}
    \includegraphics[height=5cm]{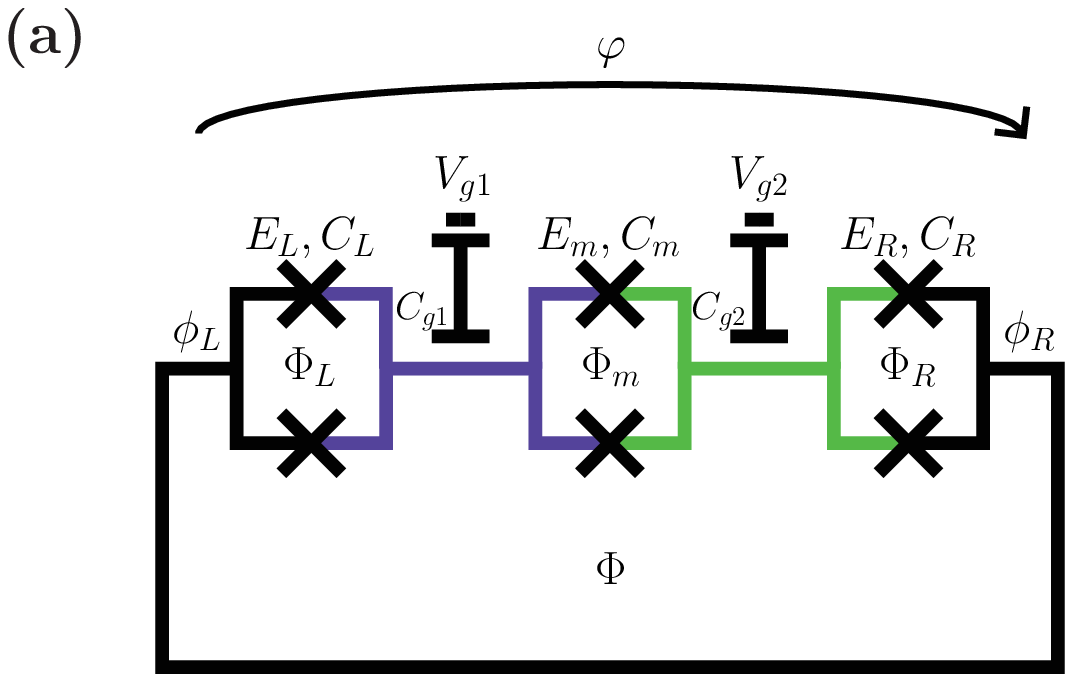} 
     \hspace{1cm}
    \includegraphics[scale=0.7]{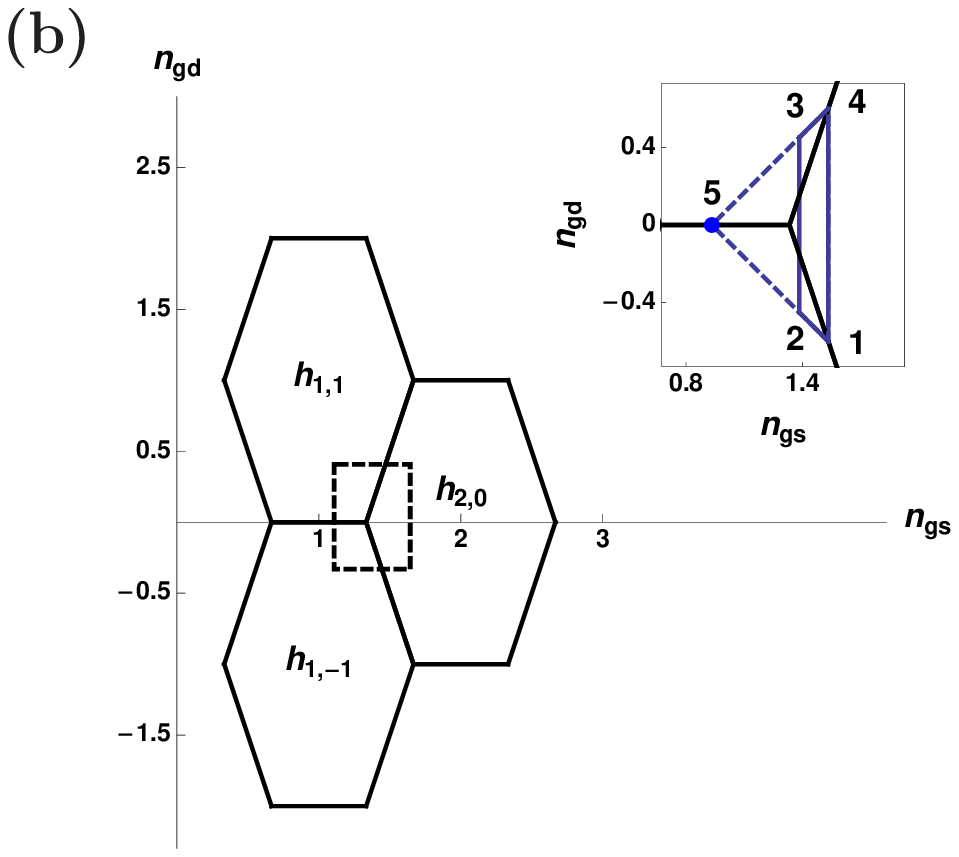}
    \end{center}
\caption{(Color online) (a) Scheme for the superconducting geometric qubit. The gate voltage and the capacitance of the $i$th island are denoted by $V_{g_i} $ and $C_{g_i}$, respectively. The $i$th SQUID is described by Josephson energy $E_i$, capacitance $C_i$, and flux $\Phi_i$. The phase across the device is $\varphi$. (b) 
The hexagonal (honeycomb) stability diagram in the $n_{gs}-n_{gd}$ plane with all the SQUIDs closed ($E_i=0$). In the inset, the loops that generate the logical gates are provided.
The trapezoidal loop (solid line $1  \rightarrow 2  \rightarrow 3 \rightarrow 4 \rightarrow 1$) generates the Hadamard gate and triangular loop  (dashed line $1  \rightarrow 4  \rightarrow 5 \rightarrow 1$) generates the $U_2$ tranformation.
	The circle represents the point in which the SQUIDs are closed and the phase across the device is reversed from $\varphi$ to $-\varphi$ in the implementation of the $U_2$ tranformation. 
	Notice that away from the degeneracy lines  the SQUIDs are open ($E_i \neq 0$) and the loops represent the projection on the $n_{gs}-n_{gd}$ plane of the degeneracy curves in the whole $\{n_{gs}, n_{gd}, E_i \}$ space. 
Here, we have chosen $k=1/3$ and a big energy gap $\alpha=\tilde{E}_C$ to visualize better the loop.
 }
	\label{fig:scheme}
\end{figure*}

\section{Physical Realization}
\label{sec:Hphys}

The physical set-up considered for the implementation of GQC is shown in Fig.~\ref{fig:scheme}(a).
It is composed of three superconducting quantum interference devices (SQUIDs) in series with two superconducting islands between them \cite{pirkkalainen10}. The SQUIDs are operated as tunable Josephson junctions which can be closed (Josephson energy $E_i=0$) and opened ($E_i \neq 0$) by controlling the magnetic flux through them. The phase difference of the order parameter across the whole device, $\varphi=\phi_R-\phi_L$, is controlled by the magnetic flux $\Phi$ through the outer-most loop.
A slightly simpler device has been already used to measure  experimentally the Berry phase \cite{mottonen08}. 

The total Hamiltonian $\hat{H}_{\mathrm{phys}}$ is composed of the charging Hamiltonian, $\hat{H}_{\mathrm{ch}} $, and the Josephson Hamiltonian, $\hat{H}_J$. The first one reads
\begin{align}
\hat{H}_{\mathrm{ch}} = & E_{C_1} (\hat{n}_1 - n_{g_1})^2 + E_{C_2} (\hat{n}_2 - n_{g_2})^2 \nonumber \\ + & E_m (\hat{n}_1 - n_{g_1}) (\hat{n}_2 - n_{g_2}),
\label{eq:H_charge1}
\end{align}
where $\hat{n}_i$ is the operator for the excess number of Cooper pairs on the island $i$ and $n_{g_i}= C_{g_i} V_{g_i} /(2e)$ is the corresponding gate charge. The charging energies are $E_{C_1} = 2 e^2 C_{\small{\sum}_2}/C^2$, $E_{C_2} = 2 e^2 C_{\small{\sum}_1}/C^2$, and $E_m = 4 e^2 C_m/C^2$. Here, $C_{\small{\sum}_i}$ is the total capacitance of the island  $i$, namely, $C_{\small{\sum}_1} = C_L+C_m+C_{g_1}+C_1^0$ and $C_{\small{\sum}_2} = C_R+C_m+C_{g_2}+C_2^0$, and $C^2 = C_{\small{\sum}_1} C_{\small{\sum}_2} - C_m^2$. Above, the $C_k^0$ is the self-capacitance of island $k$.

The Josephson Hamiltonian is given by
\begin{align}
\hat{H}_J =-\frac{1}{2} &\sum_{n_1,n_2=-\infty}^{\infty} \Big(  E_L(\Phi_L) e^{i \varphi/2} |n_1+1,n_2 \rangle \langle n_1,n_2| \nonumber \\
+ & E_m(\Phi_m)  |n_1+1,n_2-1 \rangle \langle n_1,n_2| \nonumber \\
+ & E_R(\Phi_R) e^{-i \varphi/2}   |n_1,n_2+1 \rangle \langle n_1,n_2| + \mathrm{h.c.}  \Big),
\label{eq:H_J1}
\end{align}
where $|n_1,n_2 \rangle$ denotes the eigenstate of $\hat{n}_1$ and $\hat{n}_2$ and $\{ E_i(\Phi_i) \}$ are the tunable Josephson energies which are controlled by the fluxes through the SQUIDs $\{ \Phi_i \}$.
Above we have neglected the contributions arising from the finite loop inductance and suppose that the SQUIDs are perfectly symmetric.

It is convenient to use the total charge on the two islands, $\hat{n}=\hat{n}_1+\hat{n}_2$, and the charge asymmetry, $\hat{m}=\hat{n}_1-\hat{n}_2$ as quantum numbers. Thus, the quantum state of the system is denoted by $|n,m\rangle$ and the new gate parameters are $n_{gs}=n_{g_1}+n_{g_2}$ and $n_{gd}=n_{g_1}-n_{g_2}$.
If the charging energies of the islands are equal $E_{C}=E_{C_1}=E_{C_2}$,  the charging Hamiltonian (\ref{eq:H_charge1}) with the new notation reads~\cite{leone08}
\begin{align}
\hat{H}_{\mathrm{ch}} =  \frac{2 E_C+E_m }{4} \left[(\hat{n}-n_{gs})^2 +  \frac{2 E_C-E_m }{2 E_C+E_m} (\hat{m}-n_{gd})^2 \right],
\label{eq:H_charge}
\end{align}
and the Josephson Hamiltonian can be expressed as 
\begin{eqnarray}
\hat{H}_J &=& -\frac{1}{2}  \sum_{n,m=-\infty}^{\infty} \Big(  E_L(\Phi_L) e^{i \varphi/2}  |n+1,m+1 \rangle \langle n, m| \nonumber \\
&+ & E_m(\Phi_m)  |n,m+2 \rangle \langle n,m| \nonumber \\
&+ & E_R(\Phi_R) e^{-i \varphi/2} |n+1,m-1 \rangle \langle n,m| + \mathrm{h.c.}  \Big).
\label{eq:H_J}
\end{eqnarray}

If all the SQUIDs are closed, i.e., $E_L=E_m=E_R=0$, the conventional stability diagram with a hexagonal lattice structure in the $n_{gs}-n_{gd}$ plane is recovered as shown in Fig.~\ref{fig:scheme}(b)~\cite{leone08}.  Hexagonal cells of the lattice are denoted by $h_{n,m}$ and inside the cell the state $|n, m\rangle$ is the non-degenerate ground state.
The edges of the cells represent lines of double degeneracy and the intersection points of three cells are associated with triple degeneracies.

In the following, we restrict the evolution to the vicinity of the triple degeneracy point $\{n_{gs},n_{gd} \} = \{(3-k)/2 , 0 \}$ with $k=(2 E_C-E_m)/(2 E_C+E_m)$ at which the states $|n=1,m=1\rangle \equiv |a\rangle$, $|n=1,m=-1\rangle \equiv |0\rangle$ and $|n=2,m=0 \rangle \equiv |1\rangle$ have the same charging energy.
They correspond to $1$ excess charge on the first island and no charge on the second, no charge on the first island and $1$ in the second, and $1$ charge on the first and second island, respectively. 
\begin{figure*}
    \begin{center}   
    \includegraphics[scale=0.65]{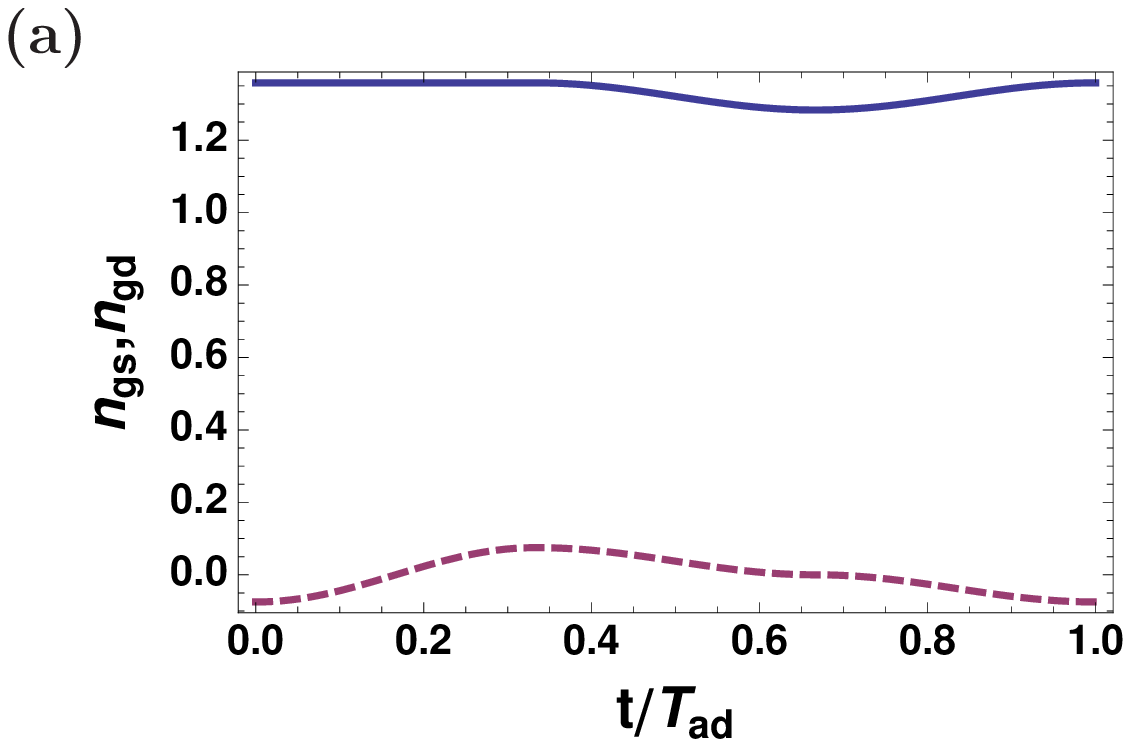} 
    \includegraphics[scale=0.65]{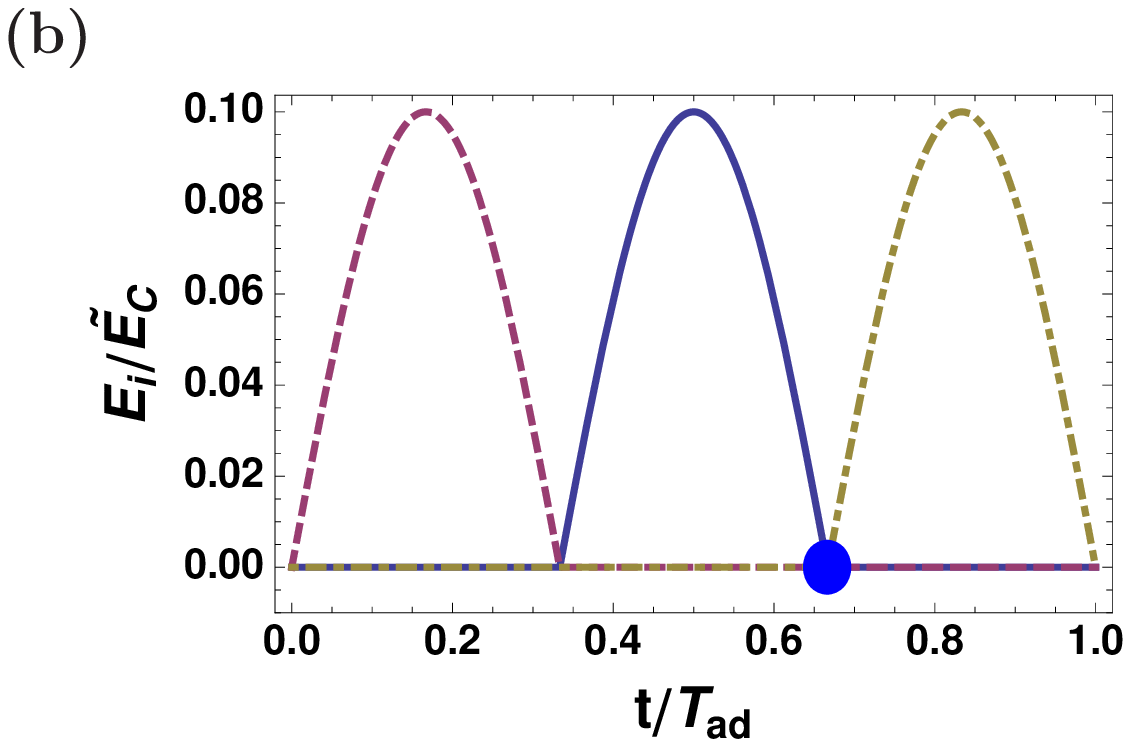}
    \end{center}
\caption{(Color online) Time evolution (normalized by the adiabatic time $T_{\rm ad}$) of the physical parameters to implement the $U_2$ transformation. (a) The gate parameters $n_{gs}$ (solid line) and $n_{gd}$ (dashed line) and (b) the SQUID parameters $|E_m|$ (solid line), $|E_L|$ (dashed line), and $|E_R|$ (dot-dashed line) in units of $\tilde{E}_C$. The blue dot denotes the point, in which the SQUIDs are closed and the phase across the device is reversed $\varphi \rightarrow -\varphi$. In all the plots the energy gap is constant $\alpha=0.1\times \tilde{E}_C$ and $k=1/3$.}
	\label{fig:loops}
\end{figure*}
For small Josephson coupling, $E_i \ll E_C$, and in the vicinity of the triple degeneracy point, the system evolution is restricted to the basis $\{ |a\rangle, |0\rangle, |1\rangle  \}$. Thus we can truncate the Hamiltonian $\hat{H}_{\mathrm{phys}}$ into this basis with the representation 
\begin{widetext}
\begin{equation}
\tilde{H}_{\mathrm{phys}}=\left(
\begin{array}{lll}
  \tilde{E}_C[(1-n_{gs})^2+k (1-n_{gd})^2] & -\frac{1}{2} E_m(\Phi_m)  & -\frac{1}{2} E_R(\Phi_R) e^{\frac{i \varphi}{2}} \\
  -\frac{1}{2} E_m(\Phi_m) & \tilde{E}_C[(1-n_{gs})^2+k (1+n_{gd})^2] & -\frac{1}{2} E_L(\Phi_L) e^{-\frac{i \varphi}{2}} \\
  -\frac{1}{2} E_R(\Phi_R) e^{-\frac{i \varphi}{2}} & -\frac{1}{2} E_L(\Phi_L) e^{\frac{i \varphi}{2}} & \tilde{E}_C[(2 - n_{gs})^2 + k n_{gd}^2]
\end{array}
\right),
\label{eq:H_phys}
\end{equation}
\end{widetext}
where $\tilde{E}_C=(2 E_C+E_m)/4$.

The mapping between the matrices $ \tilde{H}$ and $\tilde{H}_{\rm phys}$  is straightforward for the diagonal elements and the absolute values of the off-diagonal terms.
For example, we can solve the system of equations for the diagonal terms $\tilde{H}_{ii}= \tilde{H}_{\mathrm{phys},ii}$. From the first equation we obtain $ \eta=\tilde{E}_C[(1-n_{gs})^2+k (1- n_{gd})^2] -\alpha  \cos^2\theta_1 \cos^2 \theta _2$, and we solve the remaining equations to have
\begin{equation}
\left \{
\begin{array}{lll}
  n_{gs} &= & \alpha \frac{ 3 \cos^2 \theta_2 -2}{4 \tilde{E}_C}+2 (3-k)   \\
   n_{gd}&=& -\alpha  \frac{ \cos \left(2 \theta _1\right)\cos ^2 \theta _2}{4 k \tilde{E}_C}. 
\end{array}
\right.
\label{eq:n_mapping}
\end{equation}

Mapping the absolute values of the off-diagonal terms as a function of $\theta_1$ and $\theta_2$ results in
\begin{equation}
\left \{
\begin{array}{lll}
  |E_L(\Phi_L)| &=& 2 \alpha  \cos \theta_2 \sin \theta_1 \sin \theta_2 \\ 
  |E_m(\Phi_m)| &=&  2 \alpha  \cos \theta_1 \cos ^2\theta_2 \sin \theta_1  \\ 
  |E_R(\Phi_R)| &=& 2 \alpha \cos \theta_1 \cos \theta_2 \sin \theta_2 .
  \label{eq:off-diag_mapping1}
  \end{array}
\right.
\end{equation}
The parameter $\alpha$ is the energy gap of the system but it has no general  simple expression in terms of the physical parameters.
In the following, we impose that $\alpha$ is constant during the evolution and determine the corresponding evolution of the physical parameters by Eqs.~(\ref{eq:n_mapping}) and~(\ref{eq:off-diag_mapping1}).
However, the value of $\alpha$ and even its time-independence are not important for the model since the geometric operator does not depend on the value of the energy gap.
This further degree of freedom can be utilized to design a time-dependent $\alpha$ leading to ease the control of the physical parameters for the experimental implementation.

Even if Eqs.~(\ref{eq:n_mapping}) and~(\ref{eq:off-diag_mapping1}) are satisfied, comparison between Eqs.~(\ref{eq:H_general}) and (\ref{eq:H_phys}) reveals that it is not possible to match  the phases of the off-diagonal terms, and thus, a global bijective mapping does not exist.
In fact, if all the SQUIDs are open, it turns out that the two Hamiltonians have different spectra and the physical Hamiltonian has a degenerate ground state only for particular choices of the phase across the device: $\varphi=0,\pi$ (see also \cite{leone08}).
However, there is still the possibility to map the two Hamiltonians locally  if we assume that it is possible to close the SQUIDs perfectly.

\section{Single-qubit logical gates}
\label{sec:single_qubit}

Since the calculation of  the geometric unitary transformation is simpler using the model Hamiltonian $\hat{H}$, we use the following approach: first determine the loop for $\hat{H}$ which produces the desired geometric transformation and then map it into the physical parameter space.

Along a single loop it is convenient to keep $\phi_2$ constant thus restricting the curve to the $\theta_1-\theta_2$ plane.
It follows from Eqs.~(\ref{eq:connections}) and~(\ref{eq:U_gamma}) that a rectangular loop in the $\theta_1-\theta_2$ plane produces the final transformation (in the logical $\{|0\rangle,  |1\rangle\}$ basis)
\begin{equation}
  U_{\gamma} = {\rm exp} \Big [\Big (		
		\begin{array}{ll}
		 0 & e^{i \phi_2} g \\
 		-e^{-i \phi_2} g  & 0
	 \end{array}
	 \Big )	\Big ],
  \label{eq:U_single_qub}
\end{equation}
where $g= (\sin  \theta_2^M -\sin  \theta_2^m)(\theta_1^M-\theta_1^m)$, and $\theta_i^M$ ($\theta_i^m$) is the maximum (minimum) value of $\theta_i$.

\subsection{Hadamard gate}

We set $\phi_2=0$ and perform the loop 
\begin{eqnarray}
(\theta_1,\theta_2)&:& (0,0) \rightarrow (0,\pi/6)    \rightarrow (\pi/2,\pi/6)  \nonumber \\ & & \rightarrow (\pi/2,0) \rightarrow  (0, 0),
\label{eq:Had_loop}
\end{eqnarray}
which corresponds to  $\theta_1^M=\pi/2$, $\theta_2^M=\pi/6$, and $\theta_i^m=0$.

The corresponding transformation using Eq.~(\ref{eq:U_single_qub}) is the Hadamard gate (apart from an irrelevant global phase),
\begin{equation}
U_1=
 \frac{1}{\sqrt{2}}\left(
\begin{array}{cc}
 1 & -1 \\
 1 & 1
\end{array}
\right).
\label{eq:Hadamard}
\end{equation}

In this case, the mapping to the physical Hamiltonian is simple. If we set the phase $\varphi$ across the device to zero, 
the Hamiltonian corresponding to Eq.~(\ref{eq:H_phys}) can be globally mapped into the one in Eq.~(\ref{eq:H_general}) with the help of Eqs. (\ref{eq:n_mapping}) and (\ref{eq:off-diag_mapping1}).
The corresponding loop in the $n_{gs}-n_{gd}$ plane is shown in the inset of Fig.~\ref{fig:scheme}(b).

\subsection{Phase gate}
\label{sec:swap-phase}

Let us consider the case in which the model Hamiltonian (\ref{eq:H_general}) undergoes a loop with  time-independent $\phi_2 \neq 0 $.
We choose $\theta_i^M=\pi/2$, $\theta_i^M=\pi/2$ and the loop
\begin{eqnarray}
(\theta_1,\theta_2):&& (0,0) \rightarrow (\pi/2,0) \rightarrow (\pi/2,\pi/2) \nonumber \\ &&\rightarrow  (0,\pi/2) \rightarrow (0,0),
\label{eq:swap_loop}
\end{eqnarray}
which produces the transformation 
\begin{equation}
U_{2}(\phi_2)=
 \left(
\begin{array}{ll}
 0 & -e^{i \phi_2} \\
 e^{-i \phi_2} & 0
\end{array}
\right).
\label{eq:U_swap}
\end{equation}

The choice of this particular loop is due to the simplicity in calculating the unitary operator.
In fact, the path $(\pi/2,\pi/2) \rightarrow  (0,\pi/2)$ is not necessary since all the SQUIDs are closed ($E_j=0$) and the $n_{gs}$ and $n_{gd}$ are constant.
The Hamiltonian (\ref{eq:H_general}) does not change along this path [see Eqs.~(\ref{eq:n_mapping}) and~(\ref{eq:off-diag_mapping1})] which therefore is physically irrelevant.

To locally map the physical Hamiltonian $\tilde{H}_{\mathrm{phys}}$ to $\tilde{H}$  along the loop in Eq.~(\ref{eq:swap_loop}), the physically irrelevant step $(\pi/2,\pi/2) \rightarrow  (0,\pi/2)$ can be neglected.
The physical evolution can be obtained from Eqs.~(\ref{eq:n_mapping}) and~(\ref{eq:off-diag_mapping1}) and it is shown in Figs.~\ref{fig:scheme}(b) and~\ref{fig:loops}.
The sequential opening of SQUIDs $R$, $L$, and $m$ in Fig. \ref{fig:loops}(b) corresponds to the three edges of the triangular loop in Fig.~\ref{fig:scheme}(b).
However, particular attention must be paid to the choice of the phase across the device $\varphi$ for the correct mapping between the Hamiltonians.

In the beginning of the loop all the SQUIDs are closed and we fix the phase to $\varphi=2 \phi_2$.
At point $(\theta_1,\theta_2)=(\pi/2,\pi/2)$, again all the SQUIDs are closed, both the Hamiltonians are diagonal and the mapping between them is trivial.
At this point [marked with a dot in Figs.~\ref{fig:scheme}(b) and \ref{fig:loops}(b)], 
no information about the phase is present, and hence we can adiabatically reverse the phase across the device $\varphi$ to $-2 \phi_2$.
During this process, an additional diagonal contribution proportional to  $\dot{\varphi}$ arises in the physical Hamiltonian due to the ac Josephson relation. It appears in the Hamiltonian as a finite bias voltage and can be treated as an effective shift in the gate charges $\delta n_{g_i}$ in Eq.~(\ref{eq:H_charge1}), and hence, it can be eliminated by an proper choice of $n_{g_i}$ allowing to restore the desired degeneracy.
After this procedure, the right SQUID is open but, since the phase through the system has been changed, it is possible to map the Hamiltonians along the third path. The final result is a mapping between the physical Hamiltonian and the model Hamiltonian with constant phase $\phi_2$. Thus, this generates the $U_{2} \left(\phi_2\right)$ transformation.

Having established the loop to build a logical $U_2$ gate, we can obtain a phase gate if we perform two sequential loops with opposite phases $\phi_2$ and $-\phi_2$. The final transformation is
\begin{equation}
U_{\rm ph}(\phi_2)=U_{2} \left(-\phi_2\right) U_{2}\left(\phi_2\right)=
- \left(
\begin{array}{ll}
 e^{-i  2 \phi_2} & 0 \\
 0 & e^{i 2 \phi_2}
\end{array}
\right),
\label{eq:U_phase}
\end{equation}
which allows us to control the relative phase between the logical states.

\subsection{Sequential application of logical gates}

To obtain the logical transformation $U_1$ and $U_{2} \left(\phi_2 \right)$ we have assumed that $\phi_2$ is constant.
However, in a sequential application of different logical gates we must change $\phi_2$ from one loop to the other.
These changes have no effect on the dynamics of the system.
In fact, they are performed when the SQUIDs are closed, $\theta_1=\theta_2=0$ and thus, from Eq.~(\ref{eq:connections}) even if $\phi_2$ changes the associated connection vanishes.

\section{Two-qubit logical gate}
\label{sec:two_qubit}

\begin{figure}
    \begin{center}
         \includegraphics[scale=.5]{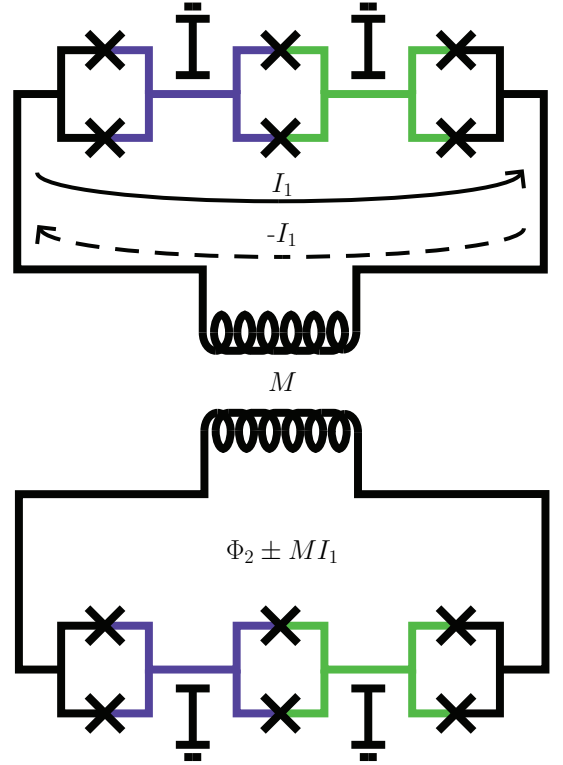} 
     \end{center}
\caption{(Color online) Circuit scheme for two qubit gate implementation. The circuits are coupled by mutual inductance, $M$. 
A current $\pm I_1$ circulating in the first circuit corresponds to an additional magnetic flux $\delta \Phi_2= \pm M I_1$  through the second. }
	\label{fig:2qubs_circuit}
\end{figure}

To complete the set of gates for universal geometric quantum computing we need a non-trivial two-qubit gate \cite{nielsen}.
To this end, we use two superconducting qubits as described above.
To couple the qubits, the mutual inductance between the two circuits is exploited as the key physical phenomenon as depicted in Fig.~\ref{fig:2qubs_circuit}.
If a current is flowing through one circuit, the flux and the overall phase across the other one are changed allowing us to implement controlled logical operations.
In the following we use the first circuit as a control qubit and the second one as a target qubit. Thus, we let the current flow through the first circuit and apply to the second qubit a geometric transformation which depends on the flux induced by mutual inductance.

\subsection{Current through the first circuit}

To couple the qubits, a stationary current is induced through the first circuit whose direction depends on the logical state of the qubit.

Denoting the Hamiltonian associated with the first circuit as $\hat{H}_{\mathrm{phys,1}}$ and the relative phase across it as   $\varphi_1$, the current operator through the first circuit is defined as $\hat{I}_1=(2 e /\hbar )\partial_{\varphi_1} \hat{H}_{\mathrm{phys,1}}$ \cite{pirkkalainen10}. 
To calculate the current at $\varphi_1=0$, we consider the variation of  the Hamiltonian for $\varphi_1\ll 1$: $\hat{H}_{\mathrm{phys},1}(0) + \delta \hat{H}_{\mathrm{phys},1}(\varphi_1)$.
The Hamiltonian $\hat{H}_{\mathrm{phys},1}(0)$ is the  dominant contribution and $\delta \hat{H}_{\mathrm{phys},1}(\varphi_1)$ is a perturbation that can be obtained as linear contribution in  $\varphi_1$.
The dominant part  $\hat{H}_{\mathrm{phys},1}(0)$ can be mapped into $\hat{H}$ and analytically diagonalized. Thus, the eigenbasis in Eq.~(\ref{eq:eigenvectors}) represent the preferred basis for our calculation.
Using the degenerate-state perturbation theory, it is possible to diagonalize  $\delta \hat{H}_{\mathrm{phys},1}(\varphi_1)$ in the degenerate subspace. The eigenstates are $|v^\prime_2\rangle = (  i |v_2\rangle +| v_3\rangle)/\sqrt{2}$ and $|v^\prime_3  \rangle= ( - i |v_2\rangle + |v_3\rangle)/\sqrt{2}$ and, in the new basis $\{ |v_1\rangle ,|v^\prime_2\rangle ,|v^\prime_3\rangle \}$, using Eq.~(\ref{eq:n_mapping}) and~(\ref{eq:off-diag_mapping1}), $\hat{H}_{\mathrm{phys},1}(\varphi_1)$ can be written as
\begin{widetext}
\begin{equation}
\tilde{H}_{\mathrm{phys},1}(\varphi_1) =\left(
\begin{array}{ccc}
 \eta+\alpha &  -f^* \varphi _1 &
   f \varphi _1  \\
  -f \varphi _1 & \eta - \alpha \cos ^2 \theta _2 \sin \left(2 \theta _1\right) \sin  \theta _2  \varphi _1 & 0 \\
 f^* \varphi _1 & 0 & \eta + \alpha  \cos ^2 \theta _2  \sin \left(2 \theta _1\right) \sin  \theta _2  \varphi _1
\end{array}
\right),
\label{eq:diagonal_H1}
\end{equation}
\end{widetext}
where $f = \alpha  \left[i \cos \left(2 \theta _1\right)+\sin \left(2 \theta_1\right) \sin \theta _2 \right ] \sin \left(2 \theta_2\right)/(2 \sqrt{2}) $.

Using Eq.~(\ref{eq:diagonal_H1}), the current operator $\hat{I}_1$ for $\varphi_1=0$ can be calculated and the average current associated with the degenerate eigenstates reads $I_{1,i}=\langle v^\prime_i |\hat{I}|v^\prime_i \rangle= (-1)^{i+1}(2e \alpha /\hbar) \sin \left(2 \theta _1\right) \sin \theta _2\cos ^2\theta _2$ (with $i=2,3$).
Thus, the current flows in opposite direction depending on the system state.

However, this is not sufficient. In fact, if initially the SQUIDs are closed, the current states $|v^\prime_2\rangle=1/\sqrt{2}(i |0\rangle+ |1\rangle)$ and $|v^\prime_3\rangle=1/\sqrt{2}(-i |0\rangle+ |1\rangle)$ do not coincide with the logical states.
Since we want the logical states to be associated with the flowing current, we must rotate the logical basis into the current state basis before opening the SQUIDs using the geometric operators $U_{\rm ph}(\pi/2)U_1$.
After this rotation, the opening of the SQUIDs could itself produce a mixing of the $|v^\prime_2\rangle$ and $|v^\prime_3\rangle$ states.
To avoid this we open the SQUIDs following the path $(\theta_1, \theta_2): (0, 0) \rightarrow (0, \pi/4) \rightarrow (\pi/4, \pi/4)$.
Using equation similar to Eq.~(\ref{eq:U_gamma}), by direct calculation, it can be verified that the current states $|v^\prime_2\rangle$ and $|v^\prime_3\rangle$ are not mixed along this path \cite{currentnote}.
Thus, the SQUIDs are open, the current flowing through the first  circuit is $I_1= \pm e \alpha /(\hbar \sqrt{2})$ depending on the initial logical state $|0\rangle$ or $|1\rangle$.

Equation~(\ref{eq:diagonal_H1}) allows us to estimate the effect of self-inductance $L$ on the system.
Even if the phase across the device is initially set to zero, the current $I_1$ flowing in the circuit induces a change $\delta \Phi_1$ in the magnetic flux. Thus, the system experiences a small additional phase $\delta \varphi_1 = \pi \delta \Phi_1/\Phi_0 = \pm \pi L |I_1| /\Phi_0$ associated with the eigenstates  $|v^\prime_2\rangle$ and $|v^\prime_3\rangle$.
As observed from Eq.~(\ref{eq:diagonal_H1}), because of the different sign of the phase shift, both the degenerate eigenstates have the same energy and the degeneracy is preserved up to the linear order in $\delta \varphi_1$.

\subsection{Two-qubit interaction Hamiltonian}

Let us assume for simplicity that the phase across the second circuit, $\varphi_2$, is initially fixed to zero. Due to the current through the first circuit, the second one experiences an additional flux through the outermost loop $\delta \Phi_2=\pm M I_1$, where $M$ is the mutual inductance between the loops.
This induces a phase increment $\delta \varphi_2= \pi \delta \Phi_2/\Phi_0 = \pm \pi M |I_1| /\Phi_0 $ depending on the initial logical state of the first qubit.

The total Hamiltonian is $\hat{H}_{\mathrm{tot}}= \hat{H}_{\mathrm{phys},1}(\varphi_1=0) + \hat{H}_{\mathrm{phys},2}(\delta \varphi_2)$.
We suppose that the quantum state of the first circuit is a combination of the degenerate states $|v_2^\prime \rangle$ and $ |v_3^\prime \rangle  $. In this subspace, the current operator $\hat{I}_1$ is diagonal and we can write the phase operator $\delta \hat{\varphi}_2$ as 
\begin{equation}
 \delta \hat{\varphi}_2= -\frac{\pi M |I_1|}{\Phi_0} (|v_2^\prime\rangle \langle v_2^\prime |-|v_3^\prime\rangle \langle v_3^\prime |).
\end{equation}

In the second circuit we suppose to have always at least one SQUID closed and thus, no current flows through it and the first circuit does not experience an additional mutual flux.
The charging Hamiltonian of the second circuit does not depend on the phase operator $\delta \hat{\varphi}_2$.
For this reason, the interaction between the two circuits occurs through the Josephson Hamiltonian of the second circuit $\hat{H}_{J,2}^{\mathrm{int}}(\delta \hat{\varphi}_2)$. Using Eq.~(\ref{eq:H_J}) it can be written as 
\begin{eqnarray}
\hat{H}_{J,2}^{\mathrm{int}}(\delta \hat{\varphi}_2)& =& -\frac{1}{2}  \sum_{n,m} \Big(  E_{L,2}(\Phi_{L,2}) e^{i \frac{\delta \hat{\varphi}_2}{2}}  \otimes |n+1,m+1 \rangle \langle n, m| \nonumber \\
& &+ E_{m,2}(\Phi_{m,2}) \openone \otimes |n,m+2 \rangle \langle n,m| \nonumber \\
& &+ E_{R,2}(\Phi_{R,2}) e^{-i \frac{\delta \hat{\varphi}_2}{2}} \otimes |n+1,m-1 \rangle \langle n,m|  \nonumber \\ 
& &+ \mathrm{h.c.}  \Big),
\label{eq:H_J_2qubs}
\end{eqnarray}
where $E_{k,2}$ and $\Phi_{k,2}$ indicate the Josephson energy and the flux of the $k$th SQUID of the second circuit, respectively, and $\delta \hat{\varphi}_2$ acts on the Hilbert space associated with the first circuit.

Writing the Hamiltonian in the degenerate subspace $\{ |v_2^\prime\rangle, |v_3^\prime\rangle \}$, we have
\begin{eqnarray}
\hat{H}_{J,2}^{\mathrm{int}}(\delta \varphi_2) &=& |v_2^\prime\rangle \langle v_2^\prime | \otimes \hat{H}_{J,2}(\delta \varphi_2)  \\ \nonumber
	&+&|v_3^\prime\rangle \langle v_3^\prime | \otimes \hat{H}_{J,2}(-\delta \varphi_2),
\label{eq:H_J_2qubs_final}
\end{eqnarray}
where $\hat{H}_{J,2}(\pm \delta \varphi_2)$ is now in the form of Eq.~(\ref{eq:H_J}).
This Hamiltonian allows us to perform an operation on the second qubit depending on the state of the first one.
In fact, if the first qubit is in the state $|v_2^\prime\rangle$ ($|v_3^\prime\rangle$), we act on the second by the Hamiltonian $ \hat{H}_{J,2}(\delta \varphi_2)$ [$ \hat{H}_{J,2}(-\delta \varphi_2)$].

  
\subsection{Two-qubit logical gate}

To be more specific, we choose to implement a particular controlled phase gate on the second qubit.
The SQUIDs are initially closed and the degenerate space is spanned by the logical states $\{ |00\rangle, |01 \rangle,|10 \rangle,|11 \rangle \}$.
We rotate the states of the first qubit by the transformation $U_{\rm ph}(\pi/2)U_1$.
In this way, when the SQUIDs are open the current flows through circuit depending on the initial logical state (for example, for $|0 \rangle$ we have current $I_1$ and for $|1 \rangle$ we have current $-I_1$).
The total phase across the second circuit is $\pm \delta \varphi_2$ and we apply a phase gate discussed in Sec. \ref{sec:swap-phase}. Then, we close the SQUIDs of the first circuit and apply a transformation $[U_{\rm ph}(\pi/2)U_1]^\dagger$ back to the logical basis.
At the end of this procedure, the logical states acquire a phase proportional to  $\delta \varphi_2$
\begin{equation}
\begin{array}{ccr}
  |00\rangle & \rightarrow   & e^{i  \delta \varphi_2} |00\rangle  ,\\
  |10\rangle & \rightarrow   & e^{-i  \delta \varphi_2} |10\rangle  ,\\
  |01\rangle & \rightarrow   & e^{-i   \delta \varphi_2}  |01\rangle , \\
  |11\rangle & \rightarrow   & e^{i \delta \varphi_2} |11\rangle.
\end{array}
\end{equation}

The total transformation can be written as 
\begin{eqnarray}
 U_{2 {\rm qubs}} &=& e^{ i \delta \varphi_2 \sigma_z} \otimes |0\rangle \langle 0| + e^{-i \delta \varphi_2 \sigma_z} \otimes |1\rangle \langle 1| \nonumber\\ 
 &=& (e^{ i \delta \varphi_2 \sigma_z} \otimes \openone )  \nonumber\\  
&& \times(\openone \otimes |0\rangle \langle 0| + e^{-2 i \delta \varphi_2 \sigma_z} \otimes |1\rangle \langle 1|),
\label{eq:two-qubsU}
\end{eqnarray}
where $ \sigma_z = |0\rangle \langle 0|- |1\rangle \langle 1|$.

From the last line in Eq.~(\ref{eq:two-qubsU}), we observe that the operator represents a controlled phase operation.
An arbitrary controlled phase gate can be produced increasing the current $I_1$ in the first circuit or by iterative application of the presented logical gate. The additional single qubit $\sigma_z$ rotation can be compensated after the iteration.

\section{Measurement scheme}
\label{sec:measurement}

The logic information is encoded in the charge degree of freedom of the two islands and the logical states correspond to $|n_1 =0 , n_2=1 \rangle = |0\rangle$ and  $|n_1 =1 , n_2=1 \rangle = |1\rangle$.
Thus the measurement of the charge on the left island is sufficient to distinguish between the two logical states.
This measurement can be done in several ways, for example, by using a radio frequency single-electron transistor (rf-SET) \cite{schoelkopf98, aassime01, bylander05} that is capacitevely coupled to the left island. The SET is embedded in a tank circuit and the readout of the charge state is performed by monitoring the reflected radio frequency signal from the circuit sensitive to the charge state on the island.

This measurement scheme has been proved to be fast and sensitive~\cite{aassime01}.
An additional advantage is that the measurement apparatus is switched off when no radio frequency signal is induced in the resonant circuit.
This point is critically important in our case since the interaction between the system and an external measurement apparatus can lead to the breaking of  the ground state degeneracy.
The effect of the additional measurement circuit on the island increases its total capacitance and it can be taken into account by a compensation in the gate voltages.

\section{Conclusions}
\label{sec:conclusions}

In this paper, we have presented means to implement ground-state geometric quantum computing in superconducting systems.
We have shown how to build all the logical gates necessary for a geometric quantum computing by manipulation of a few experimental parameters.
The main advantage with respect to the previous proposals  for geometric quantum computing is that the evolution occurs completely in the ground state.
This feature renders the implementation robust against low-temperature environments which motivates for an experimental verification.
In fact, the physical system discussed here can be potentially built and similar measurements on a simpler system have been already carried out~\cite{mottonen08}.

Although we focused on the implementation in superconducting circuits, it is possible that similar Hamiltonians 
can be found in other physical systems as happened for the first experimental proposal in Refs.~\cite{unanyan99,duan01}.

The future challenge is to study in detail the robustness of our scheme against environmental effects.
In this context, the results of Ref.~\cite{pekola09, solinas10,salmilehto10} seem promising since the non-degenerate ground state evolution has proven to be robust against low-temperature environmental noise.
If similar results are obtained for the present model, it could become a new paradigm for the implementation of robust 
geometric quantum computing.

We acknowledge Academy of Finland and Emil Aaltonen Foundation for financial support. We have received funding from the European Community's Seventh Framework Programme under Grant Agreement No. 238345 (GEOMDISS).



\begin{thebibliography}{99}
\bibitem{currentnote} The connection corresponding to the path $(\theta_1, \theta_2): (0, 0) \rightarrow (0, \pi/4) \rightarrow (\pi/4, \pi/4)$ has a diagonal contribution. However, this additional contribution can be cancelled if, after the gate application on the second qubit, we close the SQUIDs traversing the same path in the opposite direction.
\end{thebibliography}
\end{document}